\documentclass[conference]{IEEEtran}
\usepackage{amsmath}
\usepackage{graphicx}
\usepackage[noend]{distribalgo}
\usepackage{algorithm}
\usepackage{amssymb}
\usepackage{color}
\usepackage[draft]{fixme}
\usepackage{multirow}
\usepackage{afterpage}
\usepackage{soul}
\newcommand{\sizefactor}{0.9}

\definecolor{dark}{gray}{.6}

\ifCLASSINFOpdf
\else
\fi
\begin{document}
\title{Optimistic Parallel State-Machine Replication}

\author{
\IEEEauthorblockN{Parisa Jalili Marandi}
\IEEEauthorblockA{University of Lugano\\
Switzerland}
\and
\IEEEauthorblockN{Fernando Pedone}
\IEEEauthorblockA{University of Lugano\\
Switzerland}
}

\maketitle

\begin{abstract}
State-machine replication, a fundamental approach to fault tolerance, requires replicas to execute commands deterministically, which usually results in sequential execution of commands.
Sequential execution limits performance and underuses servers, which are increasingly parallel (i.e., multicore).
To narrow the gap between state-machine replication requirements and the characteristics of modern servers, researchers have recently come up with alternative execution models.
This paper surveys existing approaches to parallel state-machine replication and proposes a novel optimistic protocol that inherits the scalable features of previous techniques.
Using a replicated B+-tree service, we demonstrate in the paper that our protocol outperforms the most efficient techniques by a factor of 2.4 times. 

\end{abstract}

\IEEEpeerreviewmaketitle

\section{Introduction}
\label{sec:intro}

State-machine replication (SMR) is a fundamental approach to designing fault-tolerant services~\cite{Lam78, Sch90}. 
In this technique, to preserve consistency replicas of a service execute a unique ordered sequence of commands deterministically. 
If some of the replicas fail, the service remains available to clients through the operational replicas. 
Determinism prevents replicas from exploiting multithreading and is contrary to the nature of modern servers, which are essentially parallel (i.e., multicore architectures).
Therefore, services replicated with the state-machine approach cannot benefit from parallelism. 

Requiring replicas to execute commands sequentially limits performance.
This limitation is acknowledged by the dependability community and some approaches have been proposed to enable multithreaded replicas in state-machine replication---we survey these techniques in Section~\ref{sec:survey}.
One prominent approach is to exploit application semantics~\cite{KD2004,KWQCAD2012,p-smr}. 
The idea is to allow independent commands to execute in parallel and serialize the execution of dependent commands. 
Independent commands are those that access disjoint sections of the replica's state and therefore do not interfere with each other~\cite{Sch90}. 
Dependent commands are those that access and modify common sections of the state; executing dependent commands concurrently may result in unpredictable and inconsistent states across replicas.

This paper builds on Parallel State-Machine Replication (P-SMR)~\cite{p-smr}, a scalable multithreaded model for replication. 
Its scalability stems from the absence of any centralized component in the execution path of independent commands.
P-SMR replaces atomic broadcast, typically used in state-machine replication to order commands, by atomic multicast.
Atomic multicast creates the abstraction of groups (i.e., disjoint ordered sequences of commands), and threads in a replica can subscribe to different groups.
Clients multicast commands to one or more groups, where the groups are chosen using a deterministic mapping that depends on the command and its parameters.
The mapping is such that (a)~independent commands are likely mapped onto different groups and (b)~any two dependent commands are mapped onto at least one common group.
Consequently, independent commands multicast to disjoint groups can be executed concurrently by different threads in a replica, and dependent commands are synchronized by their common group and executed by a single thread.

Despite its highly scalable execution model, P-SMR's Achilles heel is its \emph{conservative strategy} to map commands to groups. 
Clients decide on the group that a command is multicast to based only on the command type and its parameters.
Since clients lack access to service state, they must choose groups conservatively to avoid the concurrent execution of potentially dependent commands, even if in the end these commands do not access any common service state (i.e., a \emph{false positive}).
For example, consider two commands to insert an item in a B-tree.
Since these commands may lead to common structural changes in the tree, in P-SMR they must be declared dependent, even though when executed they modify different tree nodes.
False positives protect the integrity of the service at the cost of reducing its performance with unnecessary serialization.

In this paper, we present opt-PSMR, an approach that 
replaces P-SMR's conservative strategy by a more aggressive \emph{optimistic strategy}.
In opt-PSMR, when uncertain about command interdependencies, clients identify the commands as independent. 
Replicas are augmented with additional logic to check whether concurrent execution of commands risks corrupting the replica's state (i.e., the optimistic assumption does not hold). 
If two commands deemed independent turn out to be dependent, they are multicast again using P-SMR's conservative strategy. 
%
Using a B$^+$-tree service, we demonstrate in the paper that opt-PSMR with its optimistic strategy outperforms P-SMR by a factor of 2.4 times. 

Several optimistic (and speculative) replication protocols have been proposed in the literature, typically with the goal of reducing latency (i.e., the delay between the submission of a command and the receipt of its answer).
These protocols can be broadly divided into two classes.
One class of protocols reduces latency by shortening the protocol execution when the optimistic assumption holds.
For example, when order happens spontaneously, an optimistic atomic broadcast protocol can deliver messages in fewer steps than a conservative protocol (e.g., \cite{PS98,Lam06,SousaPMO02}).
Another class of protocols reduces latency by overlapping the ordering of commands with their execution (e.g., \cite{JPPM02, KPAS99, marandi2011high}).
If the optimistic assumption does not hold, command's execution must be rolled back.
opt-PSMR differs from these protocols in that optimism is used to \emph{increase throughput} without penalizing latency.


This paper makes the following contributions: (a) it surveys parallel approaches to state-machine replication, (b) it proposes opt-PSMR, a novel approach that overcomes P-SMR's shortcomings, and (c) it assesses the performance of opt-PSMR and compares it to other replication techniques. 

The rest of the paper is structured as follows. 
Section~\ref{sec:model} describes our system model and assumptions. 
Section~\ref{sec:survey} reviews parallel approaches to SMR. 
Sections~\ref{sec:optpsmr} and~\ref{sec:evaluation} present and experimentally evaluate opt-PSMR, respectively. 
Section~\ref{sec:rwork} overviews related work and Section~\ref{sec:final} concludes the paper.

\section{System model and assumptions}
\label{sec:model}

We assume a distributed system composed of interconnected processes. There is an unbounded set $C = \{ c_1, c_2, ... \}$ of client processes and a bounded set $S = \{ s_1, s_2, ..., s_n \}$ of server processes. The system is asynchronous: there is no bound on message delays and on relative process speeds. We assume the crash failure model and exclude malicious and arbitrary behavior (e.g., no Byzantine failures). Processes are either \emph{correct}, if they never fail, or \emph{faulty}, otherwise. We assume $f$ faulty servers, out of $n = f+1$ servers. 

Processes communicate by message passing, using either one-to-one or one-to-many communication. One-to-one communication is through primitives send$(m)$ and receive$(m)$, where $m$ is a message. If sender and receiver are correct, then every message sent is eventually received. 
One-to-many communication is based on atomic multicast. Atomic multicast is defined by the primitives \emph{multicast}$(\gamma, m)$ and \emph{deliver}$(m)$, where $\gamma$ is a group of destinations.
Let relation $<$ be defined such that $m < m'$ iff there is a process that delivers $m$ before $m'$.
Atomic multicast ensures that 
(i)~if a process delivers $m$, then all correct processes in $\gamma$ deliver $m$ \emph{(agreement)}; and
(ii)~relation $<$ is acyclic \emph{(order)}.
The order property implies that if processes $p$ and $q$ deliver messages $m$ and $m'$, then they deliver them in the same order.

Atomic multicast is typically available to applications as a library (encapsulated as an agreement layer) and implemented using one-to-one communication and additional system assumptions~\cite{CT96,Lam98}. Atomic broadcast is a special case of atomic multicast where there is only one group to which all the destinations belong. 

Our consistency criterion is linearizability.
A system is linearizable if there is a way to reorder the client commands in a sequence that (i)~respects the semantics of the commands, as defined in their sequential specifications, and (ii)~respects the real-time ordering of commands across all clients~\cite{Attiya04}. 

\section{A Survey on Parallel SMR}
\label{sec:survey}
In this section we review the basics of state-machine replication and survey proposals that have adapted state-machine replication to multicore architectures. 

\subsection{Non-replicated setup}

\begin{figure*}[ht]
  \begin{center}
      \includegraphics[width=\sizefactor\textwidth]{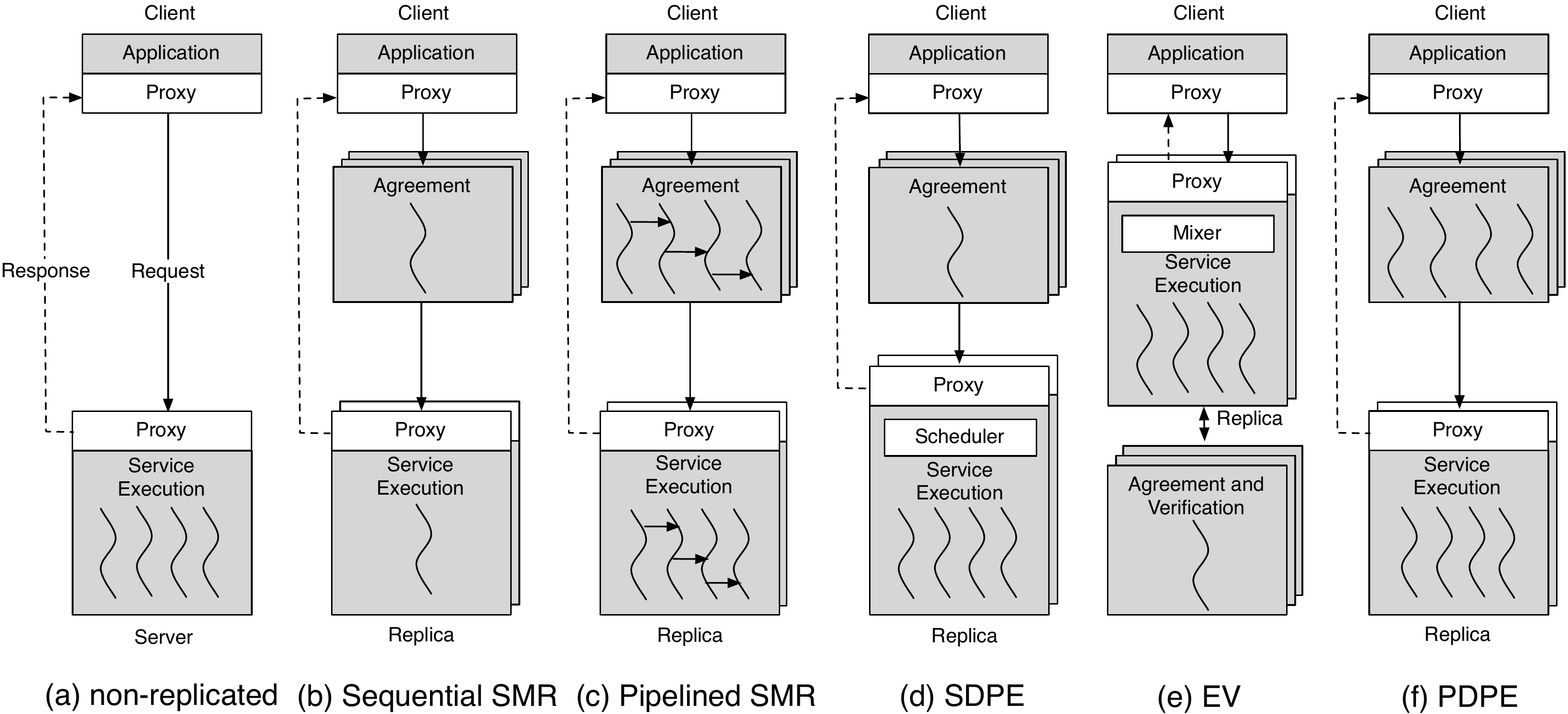} 
    \caption{Architecture differences among (a) non-replicated service, (b) sequential state-machine-replication, (c) pipelined state-machine replication, (d) sequential delivery-parallel execution (SDPE), (e) execute-verify, and (f) parallel delivery-parallel execution. Agreement layer and replicas are fault-tolerant.}
        \label{fig:architecture}
  \end{center}
\end{figure*}

A typical way for clients to interact with a (non-replicated) server is by means of \emph{remote procedure invocations}~\cite{BN84,tanenbaum1995distributed}. Clients access the service by invoking service commands with the appropriate parameters. 
Client proxies intercept client invocations and turn them into requests that include a command identifier and the marshaled parameters. 
Requests are delivered by the server proxies, which re-assemble invocations and issue them against the local service. 
Similarly to remote procedure calls, the client and client proxy (respectively, server and server proxy) can be implemented as a single process, sharing a common address space. 
The command's response follows the reverse path to the client using one-to-one communication. 
As depicted in Figure~\ref{fig:architecture}~(a), in a non-replicated service: (i) client~requests are communicated to the server directly, without passing through an agreement layer, and (ii) execution of client requests at the server can be multithreaded. 

\subsection{Sequential SMR}
\label{sec:seqsmr}



State-machine replication provides clients with the illusion of a non-replicated service, that is, replication is transparent to the clients.
A command issued by a client is handled by the client proxy, which multicasts the command to all replicas and waits for the response from one replica (see Figure~\ref{fig:architecture}~(b)). Before requests can be executed on the replicas they are ordered by the agreement layer. Since replicas execute commands deterministically and in the same order, every replica produces the same response after the execution of the same command.
%
Differently from a non-replicated service, clients remain oblivious to failures, as the service remains operational despite the failure of some of its replicas. 
In failure-free scenarios, however, a non-replicated service is often more efficient than a replicated service since in the replicated case requests reach the servers through an agreement layer and execution is single-threaded. 


\subsection{Pipelined SMR}
\label{sec:pipelinedsmr}
Having replicas execute commands sequentially by a single thread does not imply that the whole replica's logic must be single-threaded; multiple threads on a replica can cooperatively handle the requests. For example, one thread receives the requests, another executes the requests, and a third thread responds to the clients. 
In~\cite{SS2011}, the authors propose a pipelined architecture to exploit the processing power of multicore servers. 
The agreement layer (atomic broadcast) and the replicas are organized as a collection of modules connected through shared message queues where messages are totally ordered (see Figure~\ref{fig:architecture}~(c)). 
Although staging improves the throughput of state-machine replication, there is always only one thread sequentially executing the commands. 

\subsection{Sequential Delivery-Parallel Execution (SDPE)}
\label{sec:sdpe}

Replicas in classic state-machine replication, execute all the commands sequentially by adhering to the order decided by the agreement layer. 
It has been observed that a replica can execute commands that access disjoint variables (independent commands) concurrently without jeopardizing consistency~\cite{Sch90}. 
The notion of command interdependency is application-specific and must be provided by the application developer or automatically extracted from the service code. Recently, several replication models have exploited command dependencies to parallelize the execution on replicas. 
We discuss these techniques in this and the next two sections. 

To understand the concept of dependencies among commands, consider a service composed of three objects $x$, $y$, and $z$ and assume commands $C_{x}$, $C_{y}$, $C_{z}$, $C_{xy}$, where the indices indicate the objects accessed and modified by the commands. 
Commands $C_{x}$, $C_{y}$, and $C_{z}$ access disjoint objects.
Thus, they are independent and can be executed in parallel at each replica. 
Command $C_{xy}$ depends on commands $C_{x}$ and $C_{y}$ and must be serialized with $C_{x}$ and $C_{y}$. 
$C_{xy}$ can be executed in parallel with $C_{z}$ however.





\newcommand{\rb}[1]{\raisebox{1ex}[0cm][0cm]{#1}}

\begin{table*}[ht]
\normalsize
\centering
\begin{tabular}{|l|c|c|c|c|c|c|} \hline
 						& 	& 	& 		& 	& \multicolumn{2}{c|}{PDPE} \\ \cline{6-7}	
						& \rb{Sequential SMR} & \rb{Pipelined SMR} & \rb{SDPE} &	\rb{EV}		& ~~~P-SMR~~~	& 	opt-PSMR \\ \hline\hline 
Single coordination point 		& Yes	& Yes 		& Yes		& Yes		&  \multicolumn{2}{c|}{No}		\\ \hline
Scalability					& None	& Limited		& Limited		& Limited		& \multicolumn{2}{c|}{Unlimited} \\  \hline
Order on commands			& Total	& Total		& Total		& Total		&  \multicolumn{2}{c|}{Partial}	\\ \hline
Load balancing				& None	& None		& Yes		& Yes		& \multicolumn{2}{c|}{Approximative}	\\ \hline 
Application semantics		& No		& No			& Yes		& Yes		& \multicolumn{2}{c|}{Yes} 		\\ \hline
Dependency tracking		& No		& No			& Server-side	& Server-side	& \multicolumn{2}{c|}{Client-side}	\\ \hline
Execution strategy			& Conservative	& Conservative & Conservative & Optimistic & Conservative	& ~Optimistic~\, \\ \hline
Rollback					& No		& No			& No			& Yes		& No    		& Possibly 	\\ \hline
\end{tabular}
\vspace {5mm}
\caption{A comparison among approaches to parallelizing state-machine replication.}
\vspace{-5mm}
\label{table:survey}
\end{table*}%
%

To benefit from command inter-dependencies to parallelize execution, some proposals add a deterministic scheduler (also known as parallelizer) to the replicas~\cite{KD2004}. 
The scheduler delivers all the commands ordered through the agreement layer, examines command dependencies, and distributes them among a pool of worker threads for execution (see Figure~\ref{fig:architecture}~(d)). 
To distribute the commands among threads, besides considering dependencies, the scheduler can also balance the load among threads. 
Threads that are less occupied can be given more commands to execute if their execution does not conflict with the commands that are being executed by other threads. 

Although thanks to the scheduler the execution is parallelized, the scheduler delivers and dispatches commands sequentially, which restrains the overall performance from scaling.
For this reason, we identify these techniques as Sequential Delivery-Parallel Execution (SDPE). 
Adapting a sequential policy for delivery has its roots in the requirements of SMR where replicas deliver one and only one stream of ordered commands. Synchronization between the scheduler and the worker threads for dispatching commands is yet another performance overhead of this model. 


\subsection{Execute-Verify (EV)}
\label{sec:eve}
One of the shortcomings of the SDPE model is the agreement layer, where only one stream of ordered requests is generated. 
Eve addresses this issue by first executing the requests on replicas and then verifying the correctness of the states through a verification stage, hence named as Execute-Verify (EV) (see Figure~\ref{fig:architecture} (e)). 
Eve distinguishes one of the replicas as the primary to which clients send their requests. 
The primary replica organizes the requests into batches and assigns to each batch a unique sequence number. 
The primary then transmits the batched requests to the other replicas. 
All the replicas, including the primary, are equipped with a deterministic \emph{mixer}. 
Using the application semantics, the mixer converts a batch of requests to a set of parallel batches such that all the requests in a parallel batch can be executed in parallel. 
Once the execution of a parallel batch terminates, replicas calculate a token based on their current state and send their token to the verification stage. 
The verification stage investigates the equality of the tokens. 
If the tokens are equal, replicas commit the requests and respond to the clients. 
Otherwise, replicas must roll back the execution and re-execute the requests in the order determined by the primary as it was batching the requests. The verification stage also adds to Eve the advantage of detecting concurrency bugs.

Similar to the scheduler in the SDPE model, the mixer in Eve may restrict the execution performance since the content of all the requests must be scrutinized by the mixer before they can be executed. Moreover, the primary replica might be overwhelmed by the amount of requests it receives. The verification stage is another synchronization point that besides the mixer and the primary replica can threaten the scalability of this approach.

\subsection{Parallel Delivery-Parallel Execution (PEPD)}
\label{sec:pepd}



Motivated by the shortcomings of the previous models, P-SMR proposes to parallelize command delivery in addition to command execution~\cite{p-smr}; hereafter we refer to this model as Parallel Delivery-Parallel Execution (PDPE). 
P-SMR has no scheduler and several threads on replicas concurrently deliver and execute multiple disjoint streams of ordered commands. 
To preserve correctness, commands in each stream must be independent from the commands in any other stream. 
To ensure independency among the concurrently delivered streams, unlike previous approaches in which command dependencies are determined at the replicas, in P-SMR command dependencies are determined by the clients, before commands are ordered. 
Commands in P-SMR are ordered by an atomic multicast library and clients multicast independent commands to different multicast groups. 
P-SMR implements a fully parallel model in which independent commands are ordered, delivered, and executed in parallel. 
Dependent commands are ordered through dedicated multicast groups and executed sequentially, as we explain next (see Figure~\ref{fig:architecture}~(f)). 



In P-SMR clients submit commands to the client proxies, which determine the destination groups of commands based on command dependencies. 
To guarantee concurrent execution of independent commands, client proxies assign independent commands to different multicast groups and to guarantee sequential execution of dependent commands, client proxies assign at least one common group to every two dependent commands. 
The amount of concurrency in a service depends on the interdependencies among the service's commands. 
P-SMR organizes server threads into $K$ multicast groups such that the $i$-th thread of each replica, $t_i$, belongs to group $g_i$. 
A thread $t_i$ executes the commands broadcast to $g_i$ concurrent to thread $t_j$ who executes the commands broadcast to $g_j$. 
The two threads, however, must synchronize their execution if a command is multicast to both $g_i$ and $g_j$. 
To synchronize, one of the threads, chosen deterministically among the two threads, $t_i$, waits for a notification from the other thread, $t_j$. 
$t_j$ notifies $t_i$ and waits for a notification from $t_i$ to resume its execution. 
Once $t_i$ executes the common command it notifies thread $t_j$.




\subsection{Summary}
\label{subsec:summary}

Table~\ref{table:survey} shows the main differences among the techniques we have discussed in this section. 
The table also contains opt-PSMR, the approach we introduce in Section~\ref{sec:optpsmr}.

Both SDPE and EV have centralized entities that can limit scalability: the scheduler and the agreement layer in SPDE; the mixer, the primary replica, and the verification layer in EV. 
PDPE does not include central roles in its design. 
Moreover, differently from other approaches, PDPE orders requests using an atomic multicast, as opposed to an atomic broadcast.

The parallelizer in SDPE and the mixer in EV also perform load balancing on the server side. 
Although in a limited way, clients in PDPE can try to distribute the load evenly among server threads (e.g., by multicasting read commands to different groups).

SPDE, EV, and PDPE rely on tracking command dependencies to parallelize execution on replicas. 
In SDPE and EV, command dependencies are checked on the server side. 
In PDPE, however, it is the clients that track dependencies and submit commands to the appropriate multicast groups.
Determining command dependencies in P-SMR is conservative and can lead to false positives. 
In opt-PSMR, commands are handled optimistically by the client proxies and on the server side appropriate actions are taken to avoid inconsistencies. 

Due to their optimistic nature, EV and opt-PSMR may be subject to rollbacks.
In opt-PSMR, however, depending on the application, execution rollbacks might not be necessary as we show in the next section.




\section{Optimistic P-SMR (opt-PSMR)}
\label{sec:optpsmr}
In this section, we motivate the need for opt-SMR, describe the novel technique in detail, and argue about its correctness.
%
\subsection{Motivation for opt-PSMR}
\label{sec:opsmr-motiv}



Consider a B$^+$-tree service that stores key-value entries where keys are integers and values are strings, and the following operations are supported: \texttt{read(in: int k, out: char[] v)}, \texttt{update(in: int k, char[] v)}, \texttt{delete(in: int k)}, \texttt{insert(in: int k, char[] v)}, where \texttt{k} is a key, \texttt{v} is a value, and \texttt{in} and \texttt{out} specify the input and output parameters of a command respectively. 


P-SMR assumes a configurable multiprogramming level (i.e., the number of threads at each replica).
Assume there are $K$ threads per replica and the same number of groups so that the $i$-th thread at each replica is part of the $i$-th group.
Clients (actually client proxies) map commands to groups using the following strategy.
Commands to \texttt{read} and \texttt{update} key \texttt{k} are mapped onto a single group $g$ (e.g., using range partitioning, $g = (\lfloor$ \texttt{k}$K/M$ $\rfloor) + 1$, where $M$ is the value of the largest key in the key space).
%
%

As a consequence, commands assigned to different groups can execute concurrently on replicas, even if they all are \texttt{update} operations.
Commands on the same key are multicast to the same group $g$ and executed sequentially by the thread associated with $g$.
Unlike \texttt{read} and \texttt{update} operations, \texttt{insert} and \texttt{delete} may cause structural changes in the tree (i.e., splits and merges).
Structural changes can spread to many nodes of the tree and interfere with the execution of other operations. 
Since it is impossible for a client to predict these changes, to preserve correctness, clients conservatively assume that \texttt{insert} and \texttt{delete} are dependent on all the other operations and multicasts them to all the groups.
All threads in a replica deliver \texttt{insert} and \texttt{delete} commands. Upon delivering such a command, threads coordinate so that a single thread executes the command.



\begin{figure}[ht]
  \begin{center}
      \includegraphics[width=\sizefactor\columnwidth]{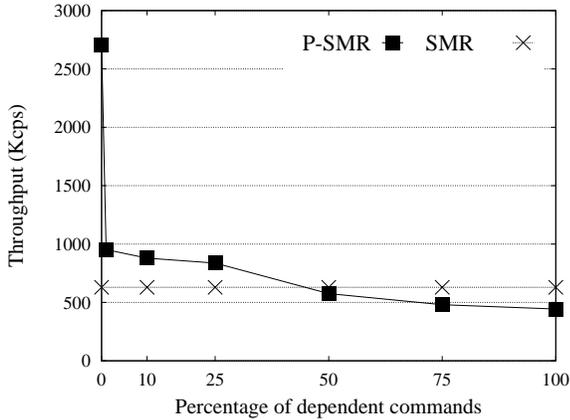} 
    \caption{Throughput of P-SMR and SMR with a workload composed of dependent and independent commands; for details of the experiment see Section~\ref{sec:evaluation}.}
        \label{fig:psmr-performance}
  \end{center}
\end{figure}

The sequential mode of a multithreaded replica in P-SMR is more expensive than the (sequential) execution mode of a single-threaded replica in traditional SMR. 
To switch to sequential mode, all the threads in P-SMR communicate and pause their execution so that only one of the threads executes dependent commands. 
In SMR, the single-threaded replica simply delivers and executes a stream of commands, without need of any synchronization operations.
Synchronization among threads has a negative impact on P-SMR's performance. 
Sequential execution of dependent commands is a must for preserving consistency and thus, the performance overhead incurred by the replica's sequential execution mode is inherent to P-SMR. 

Figure~\ref{fig:psmr-performance} compares the performance of P-SMR and SMR for the B$^+$-tree example. 
As the percentage of \texttt{insert} and \texttt{delete} operations in the workload increases (x-axis), the throughput of P-SMR reduces and gradually falls below the performance of SMR (for details of the experiment see Section~\ref{sec:evaluation}). 
With 100\% of dependent commands, threads in P-SMR must coordinate for every delivered operation.
The difference in performance between SMR and P-SMR with 100\% of dependent operations can be understood as the cost of synchronizing threads.
The break-even point of SMR and P-SMR is when slightly fewer than half of commands are dependent.
With few dependent commands in the workload, P-SMR largely outperforms SMR, since the execution is mostly concurrent.

In light of the tradeoff shown in Figure~\ref{fig:psmr-performance}, service designers should strive to reduce command interdependencies (e.g., avoiding false sharing in the service data structures).
Although interdependencies cannot be always avoided, what really limits P-SMR's performance is that clients cannot accurately tell when interdependencies happen since they do not have the service's state.
Therefore, clients \emph{conservatively} identify some commands as dependent to prevent potential inconsistencies that can arise due to their concurrent execution at the replicas. 
In the B$^+$-tree example, \texttt{insert} and \texttt{delete} operations are among such commands. Clients multicast these commands to all the groups and therefore all the threads on the replicas deliver them and collaboratively enter the sequential mode. 
Not all the \texttt{insert} and \texttt{delete} operations, however, result in changes in the tree's structure. 
This subset of commands that are categorized as dependent but could be executed in parallel are false positives. 
In fact, we can expect the number of structural changes to decrease as the tree grows.
Based on this observation, in the next section we propose a new technique to increase the concurrency of P-SMR in the presence of  dependent commands. 


\subsection{Overview of opt-PSMR}
\label{sec:optpsmr-over}
Inspired by the inefficiency of P-SMR at executing dependent commands, and based on our observation in Section~\ref{sec:opsmr-motiv}, in this section we propose an optimistic technique to increase concurrency of P-SMR.  

Considering the impact that service state can have on the command interdependencies, we differentiate between two types of interdependencies: we refer to command interdependencies as \emph{static}, if they are determined irrespective to the service state, and as \emph{dynamic}, if they are determined with respect to the service state. 
Static interdependencies are known before runtime, that is, they can be assessed upon inspecting the command's code and parameters.
For example, two updates on the same entry are statically dependent.
Dynamic interdependencies can only be known when commands are executed, based on the state of the replica. 
Two insert operations on different keys are dynamically dependent if they lead to structural changes in the tree.

We recall from Section~\ref{sec:pepd} that in P-SMR commands are mapped to multicast groups by the client proxies.
Since client proxies do not have access to the service state, which resides on the servers, client proxies cannot exploit dynamic dependencies among commands.
Client proxies handle dynamic dependencies by conservatively declaring the involved commands as dependent.
%
%
%
In opt-PSMR, clients handle dynamic dependencies by optimistically declaring the involved commands as independent.
Dynamic dependencies are tracked at the servers.
Upon detecting a dynamic dependency among commands, the replica multicasts the command again, as a dependent command.
Note that dynamic dependency tracking must be deterministic at replicas.

We define CC-G (conservative command-to-group) and OC-G (optimistic command-to-group) functions, for mapping the commands to multicast groups. 
CC-G is conservative mapping used by P-SMR: it declares two commands dependent if they have static dependencies or could possibly have dynamic dependencies.
OC-G is built on static dependencies only; commands that may have dynamic dependencies only are deemed independent.


A client~proxy in opt-PSMR calls the OC-G function to determine the groups to which commands will be optimistically multicast. 
Whenever a worker thread $t_k$ on a server proxy $s_i$ delivers an optimistically multicast command $C$, $t_k$ performs a \emph{safety check} on $C$ against the current state of the service. 
A safety check is application specific and should be provided by the application developer or automatically computed from the service code. 
During the safety check, $t_k$ seeks to figure the consequences of $C$'s execution on the state. 
Depending on the application, the changes of a command's execution on the state can either be identified without actually executing the command or during the execution of the command. 
Two cases can happen as the result of safety check:


\noindent \textbf{Fail.} If the safety check identifies that the modifications of $C$ to the service state are detrimental to the concurrent execution of other threads, $t_k$ fails $C$ and calls the CC-G function to identify the set of new groups to which the failed command must be sent. 
In this case, we say that the optimistic assumption has failed. 
If the safety check can be performed without executing $C$, the replica's state remains intact and a rollback is not needed. Otherwise, if the safety check requires executing $C$, then a fail must rollback $C$'s effects to the replica. 

\noindent \textbf{Pass.} If the safety check decides that $C$'s execution will not interfere with the concurrent execution of other commands we say that the optimistic assumption has succeeded. In this case if $C$ was not executed during the safety check, $t_k$ executes it. Otherwise, $t_k$ continues with the rest of the delivered commands.

Since a failed command passes through the agreement layer twice and it might require rollbacks, the cost of failed commands is reflected on the latency and possibly the CPU usage of the replicas, specially if rollbacks are needed. 
Thus the optimistic assumption is of practical interest if for a given execution, the ratio of fails is lower than the ratio of passes.

We illustrate opt-PSMR with the B$^+$-tree example. In P-SMR as a client~proxy is not aware of the state changes caused by the \texttt{insert} and \texttt{delete} operations, it conservatively categorizes them as dependent and multicasts them to all the multicast groups. A client proxy in opt-PSMR, however, selects the multicast groups of these operations based on the keys they access, similarly to the \texttt{read} and \texttt{update} operations. The optimistic assumption in this example is that the consequences of \texttt{insert} and \texttt{delete} operations will at most affect the node that contains the accessed key. Basically, the CC-G and OC-G functions can be defined as follows, where $K$ is the number of threads on replicas and $M$ is the value of the largest key in the key space:

\vspace{1mm}
\begin{itemize}
\small
\itemsep0.08em
\item[] \hspace{-10mm}\textbf{function} \textit{CC-G}$(cid,x)$
\item[] \hspace{-3mm}\textbf{if} \texttt{$cid \in \{$read,update$\}$} : return($\lfloor xK/M \rfloor$)
\item[] \hspace{-3mm}\textbf{else} return($\it{ALL\_GROUPS}$)
\end{itemize}
\vspace{1mm}
\begin{itemize}
\small
\itemsep0.08em
\item[] \hspace{-10mm}\textbf{function} \textit{OC-G}$(cid, x)$
\item[] \hspace{-3mm}return($\lfloor xK/M \rfloor$)
\end{itemize}
\vspace{1mm}


If a thread $t_k$ on a server proxy delivers an \texttt{insert} or \texttt{delete} operation that is multicast optimistically, it issues a safety check against the current state of the tree. If $t_k$ figures that the execution of the \texttt{insert} or \texttt{delete} will cause changes that will propagate to other nodes of the tree, it will fail the command and call the CC-G function and conservatively retry the command. Otherwise $t_k$ can successfully execute the command. We note that in the B$^+$-tree example, the potential changes of a command's execution on the tree's structure can be determined without actually executing it. Therefore rollbacks are never needed. 
\begin{algorithm}
\small
\mbox{\textbf{Algorithm 1: Optimistic P-SMR (opt-PSMR)}}
\begin{distribalgo}[1]
\vspace{-3mm}
\INDENT {\emph{A client proxy $c$ executes a call to command $C$ with \\ \hfill identifier $cid$ and $\mathit{input}$ and $\mathit{output}$ parameters as follows:}
}	\STATE $\gamma \leftarrow \textit{OC-G}(cid, \mathit{input})$
	\COMMENT{$\gamma$ is the set of groups involved in C}
	\STATE multicast$(\gamma, [\gamma, \mathit{opt}, c, cid, \mathit{input}])$
	\STATE wait for first response
	\STATE $\mathit{output} \leftarrow$ response
	\STATE return
\ENDINDENT
\vspace{1.5mm}
\INDENT{\emph{Thread $t_k$ at a server proxy $s_i$ executes a command as follows:}}
	\INDENT { \textbf{upon} deliver$([\gamma, \mathit{mode}, c, cid, \mathit{input}])$ for the first time}
		\IF{$\gamma$ is a singleton}
				\IF[is $C$ in optimistic mode?]{$\mathit{mode} = \mathit{opt}$}
					    \IF{$\textit{safety\_check}(cid, \mathit{input})$ fails}
						      \STATE $\gamma \leftarrow \textit{CC-G}(cid, \mathit{input})$
							\STATE multicast$(\gamma, [\gamma, \mathit{csv}, c, cid, \mathit{input}])$
					   \ENDIF					
				\ENDIF	
			\STATE \emph{// Thread $t_k$ is in parallel mode}
			\STATE execute $cid$ with $\mathit{input}$ parameters
			\STATE send response to $c$
		\ELSE
			\STATE \emph{// Thread $t_k$ is in synchronous mode}
			\STATE $e \leftarrow min\{ j : g_j \in \gamma\}$
			\COMMENT{pick a thread deterministically}
			\IF{$k = e$}
				\INDENT{\textbf{for each} $j \neq k$ such that $g_j \in \gamma$}
					\STATE wait for signal from $t_j$
				\ENDINDENT		
				\STATE execute $cid$ with $\mathit{input}$ parameters
				\STATE send response to $c$
				\INDENT{\textbf{for each} $j \neq k$ such that $g_j \in \gamma$}
					\STATE signal $t_j$
					\COMMENT{let thread $t_j$ resume its execution}
				\ENDINDENT		
			\ELSE
				\STATE signal $t_e$
				\STATE wait for signal from $t_e$
				\COMMENT{thread $t_k$ pauses its execution}
			\ENDIF
		\ENDIF
	\ENDINDENT
\ENDINDENT
\end{distribalgo}
\label{psmralg}
\end{algorithm}


\subsection{Algorithm in detail}
\label{sec:alg}

Algorithm~1 presents opt-PSMR in detail, for the case in which rollbacks are not needed (see Section~\ref{sec:evaluation}).
To execute command $C$, invoked by an application client (line~1 in Algorithm~1), the client proxy determines all groups $\gamma$ involved in $C$ using the OC-G function (line~2) and multicasts $C$ and its input parameters to groups in $\gamma$ (line~3). 
The client proxy then waits for the first response from the replicas (line~4), assigns the response received to the output parameters of $C$ (line~5), and returns to the application (line~6). 
Upon delivering $C$ (line~8) for the first time, if $C$ was multicast to a single group (line~9), thread $t_k$ tests whether $C$ is in optimistic mode (line~10) and whether it fails the safety check (line~11), in which case $t_k$ calls CC-G to conservatively determine the set of groups involved in $C$ (line~12) and multicasts $C$ to the groups specified by CC-G  in conservative mode (line~13).
If $C$ is in conservative mode, $t_k$ executes $C$ in parallel mode (line~15) and sends the response to client (line~16). 
If $C$ was multicast to multiple groups (line~18), then $t_i$ continues in synchronous mode and determines the thread, $t_e$, among $C$'s destinations, that will execute $C$ (line~19). 
If $t_k$ is in charge of executing $C$ (lines~21--26), it waits for a signal from every other thread in $C$'s destination set (lines~21--22), executes $C$ (line~23), sends the response to the client (line~24), and then signals all other threads in $C$'s destination set to continue their execution (lines~25--26). 
If $t_k$ is not in charge of $C$, it signals thread $t_e$ (line~28) and waits for $C$ to complete (line~29).

\begin{figure*}[t]
  \begin{center}
    \begin{tabular}{cc}
      \includegraphics[width=\sizefactor\columnwidth]{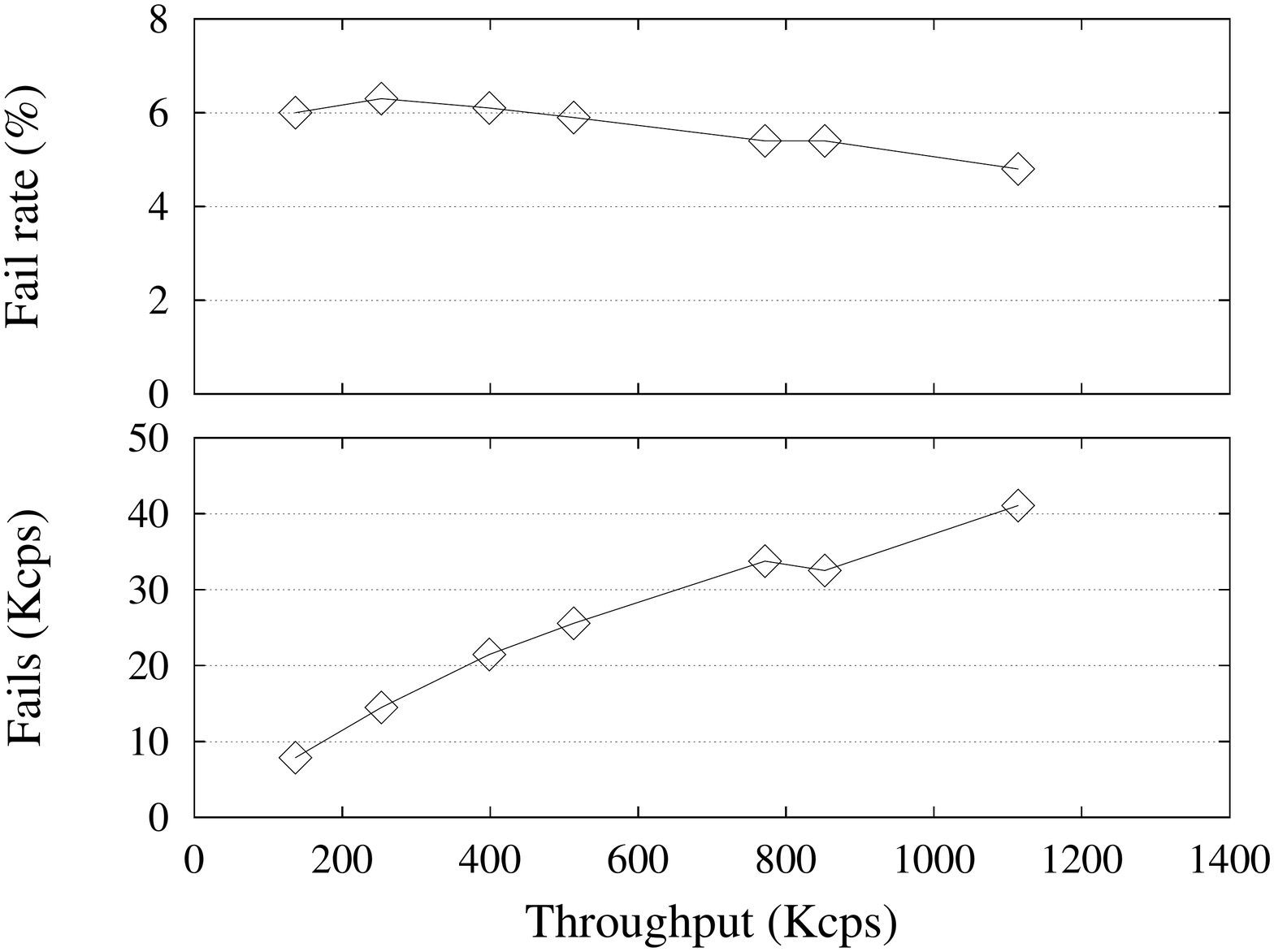}&
       \includegraphics[width=\sizefactor\columnwidth]{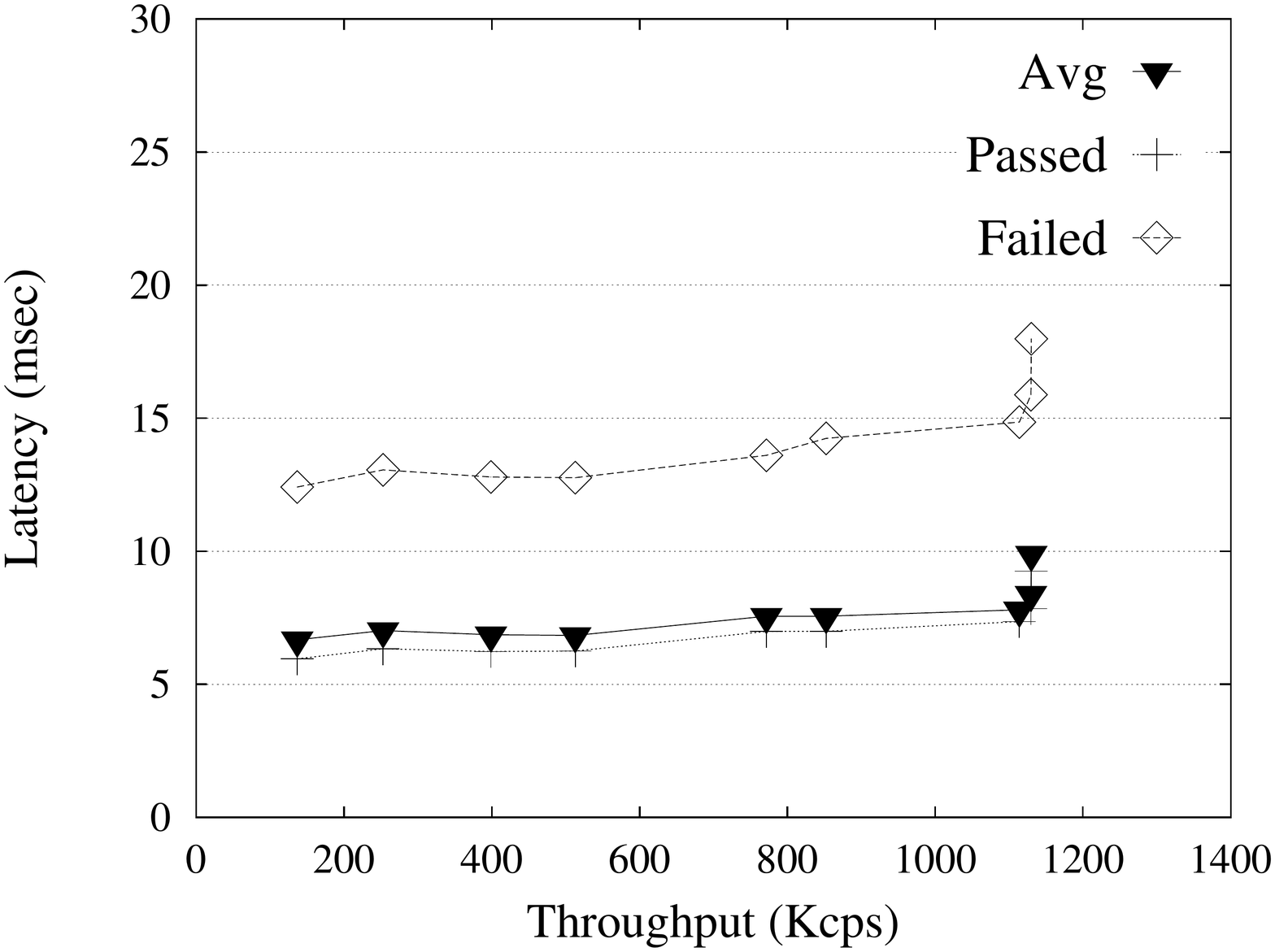}\\
    \end{tabular}
    \caption{The impact of failed commands on the latency of opt-PSMR with a dependent-only workload. The following metrics are shown: fail rate and number of failed commands versus the throughput measured in Kilo commands per second (Kcps) (left graphs); latency in milliseconds for failed, passed, and all the commands (right graph).}
    \vspace{-5mm}
    \label{fig:aborts-latency}
 \end{center}
\end{figure*}
\subsection{Correctness}
\label{sec:correct}

In this section we argue that opt-PSMR is linearizable.
From the definition of linearizability (see Section~\ref{sec:model}), there must be a permutation $\pi$ of commands in $\mathcal{E}$ that respects (i)~the real-time ordering of commands across all clients, and (ii)~the semantics of the commands. 
Our argument below relies on the fact that P-SMR is itself linearizable~\cite{p-smr}. 
opt-PSMR differs from P-SMR for commands for which the output of CC-G and OC-G functions differ.
Here we argue that opt-PSMR is linearizable for this subset of commands. 
Let $C_x$ and $C_y$ be two commands in $\mathcal{E}$ submitted by clients $c_x$ and $c_y$, respectively, where at least for one of the commands the outputs of CC-G and OC-G functions differ. There are two cases to consider. 

\noindent \emph{Case (a): For both commands $C_x$ and $C_y$, the output of CC-G and OC-G differ.} 


Before executing any of these commands a replica will perform the safety check. 
We assume thread $t_x$ performs the safety check for command $C_x$ and thread $t_y$ performs the safety check for command $C_y$. 
Three cases are possible: (1)~the safety check for both commands passes, (2)~the safety check for only one of the commands passes (without loss of generality we assume this command is $C_x$), or (3)~the safety check for both the commands fail. 
In case (1), the commands are independent and the correctness of this case directly follows from the correctness of P-SMR at executing independent commands~\cite{p-smr}. 
In case (2), only command $C_y$ is executed and according to the replicas' logic, command $C_x$ is not executed and must be multicast again with the output of the CC-G function and treated as a dependent command. 
The correctness of this case follows from the correctness of P-SMR when executing dependent commands. 
In case (3), none of the commands are executed, and therefore the state of the replica is not changed. 
These commands are then multicast again with the output of the CC-G function and similar to case (2); correctness follows from P-SMR's correctness at executing dependent commands. 


\noindent \emph{Case (b): For only one of the commands, either $C_x$ or $C_y$, the outputs of CC-G and OC-G functions differ.}

Without loss of generality, we assume the outputs of the CC-G and OC-G functions differ for command $C_x$. 
The replicas will perform the safety check for command $C_x$ only and will directly execute command $C_y$. 
If the safety check for command $C_x$ passes, commands $C_x$ and $C_y$ are independent and the correctness of this case follows directly from the correctness of P-SMR with independent commands. 
If the safety check fails, command $C_x$ is multicast again as a dependent command. 
The correctness of this case follows from the correctness of P-SMR at executing dependent commands.

\begin{figure*}[t]
  \begin{center}
    \begin{tabular}{cc}
      \includegraphics[width=\sizefactor\columnwidth]{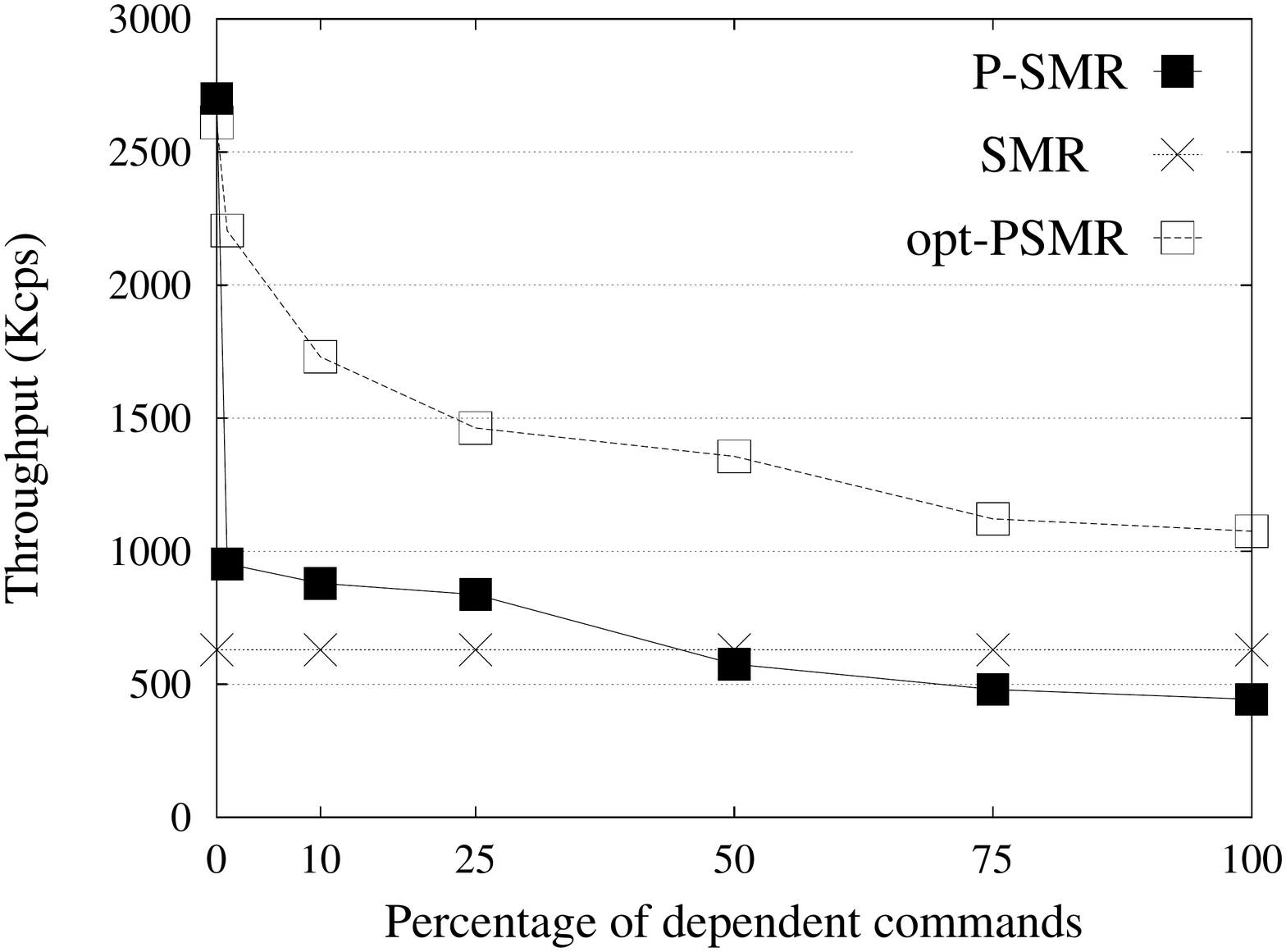}&
       \includegraphics[width=\sizefactor\columnwidth]{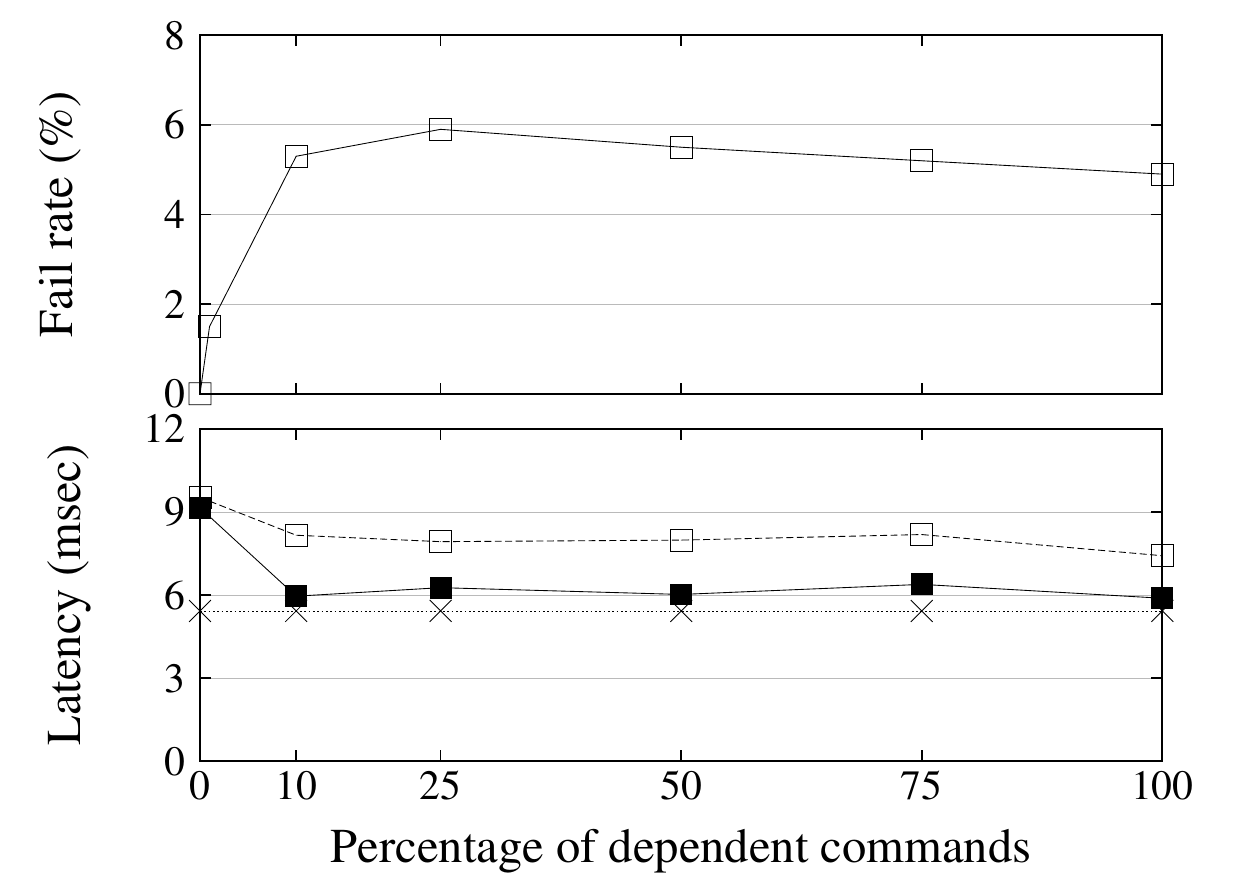}\\
    \end{tabular}
    \caption{The impact of dependent commands on the performance of SMR, P-SMR, opt-PSMR; x-axis  shows the percentage of dependent commands in the workload; the following metrics are shown: maximum throughput in Kilo commands executed per second (Kcps) (left);  average latency in milli seconds (bottom-right); the percentage of failed commands (top-right).}
    \label{fig:mixed-workloads}
 \end{center}
\end{figure*}

\begin{figure*}[ht]
  \begin{center}
    \begin{tabular}{cc}
      \includegraphics[width=\sizefactor\columnwidth]{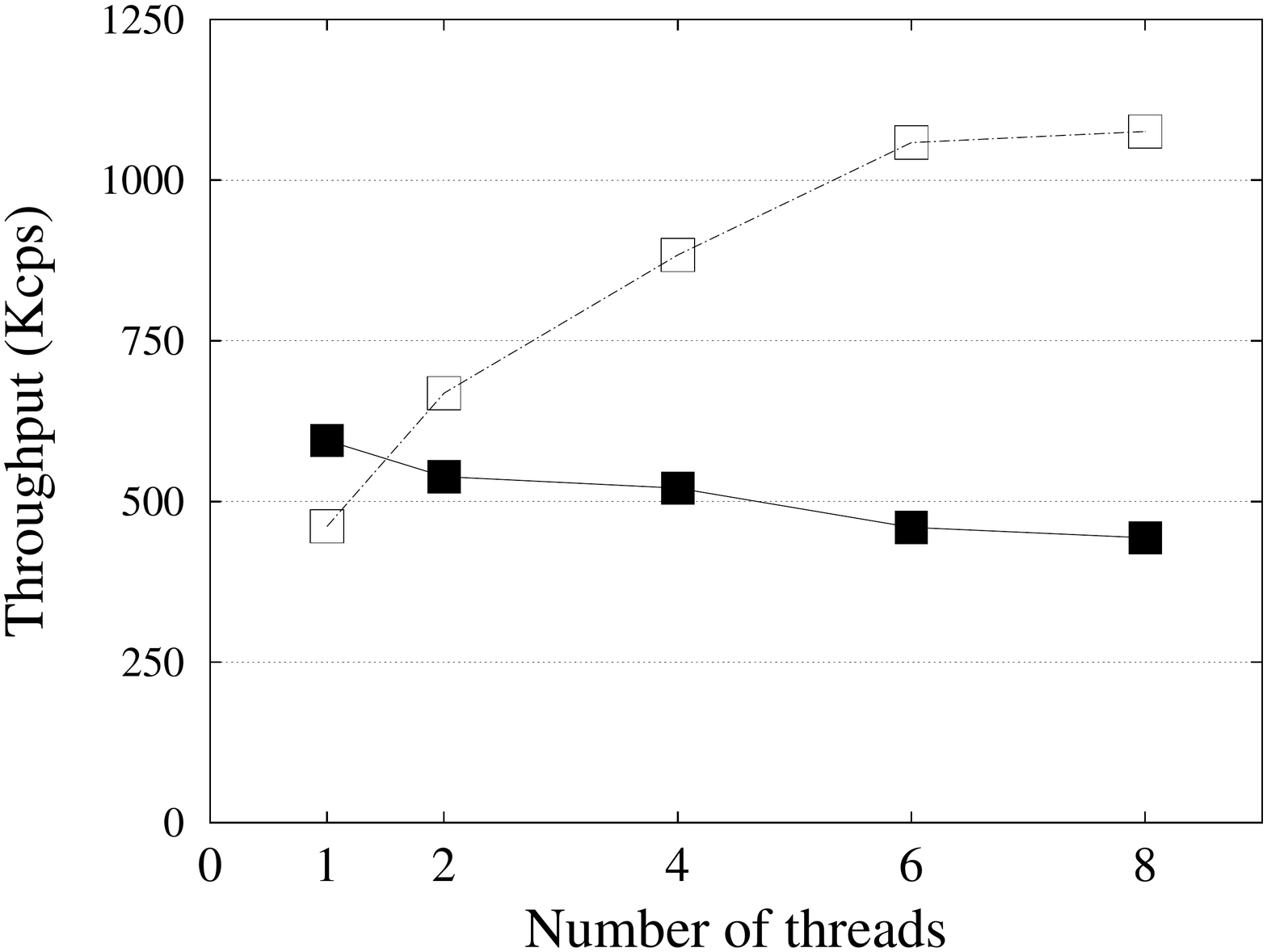}&
       \includegraphics[width=\sizefactor\columnwidth]{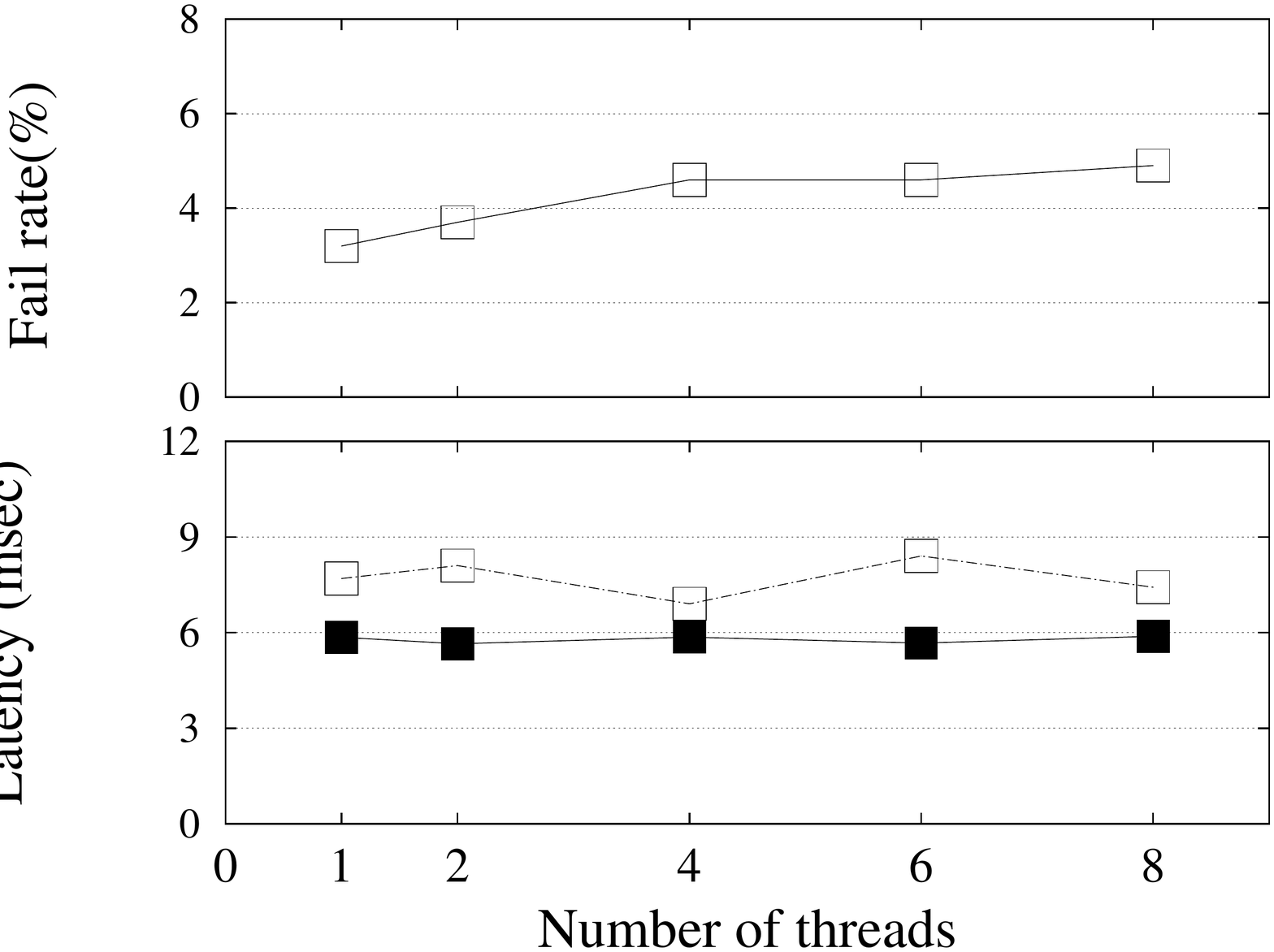}\\
      \includegraphics[width=\sizefactor\columnwidth]{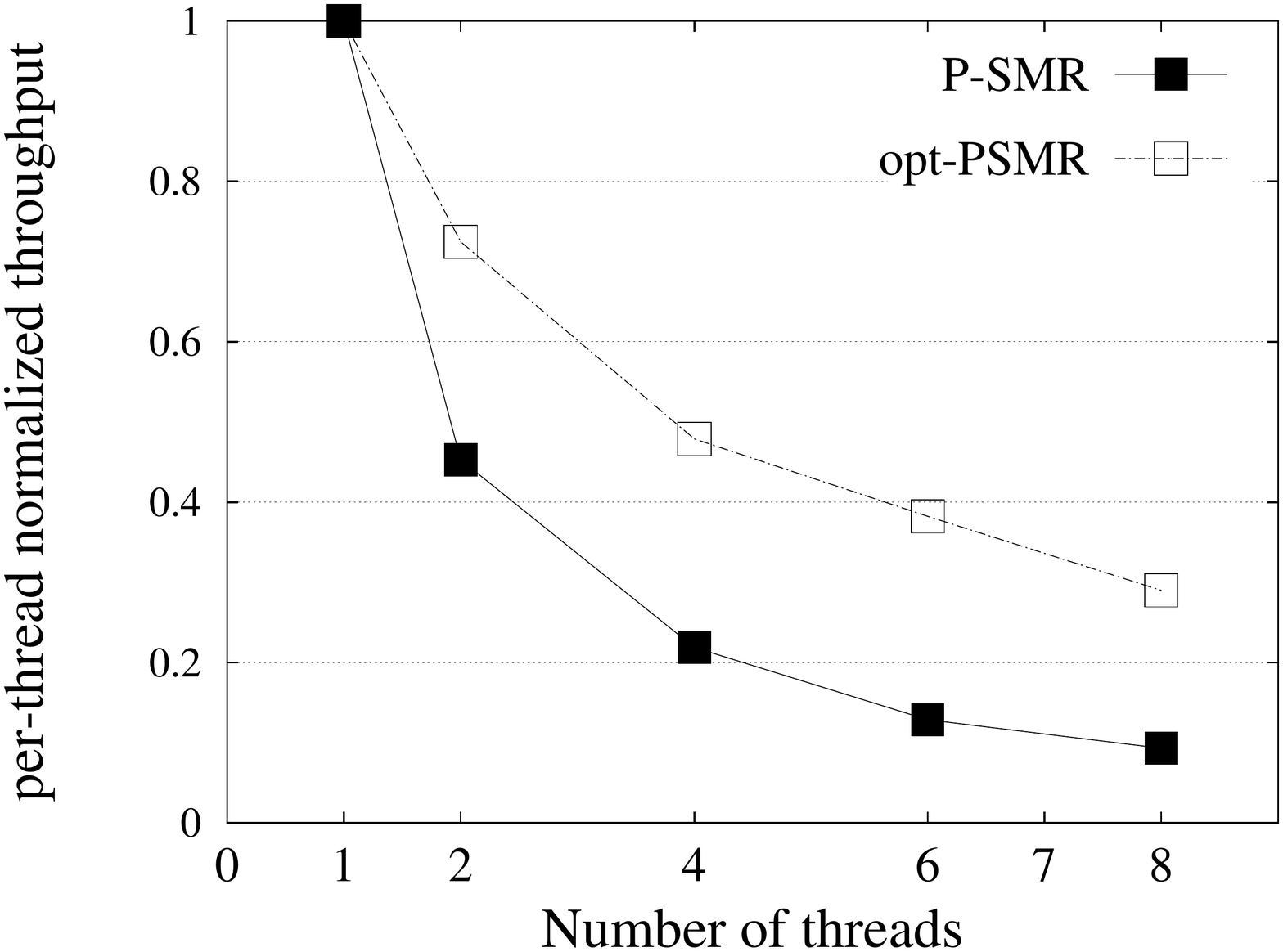}&
       \includegraphics[width=\sizefactor\columnwidth]{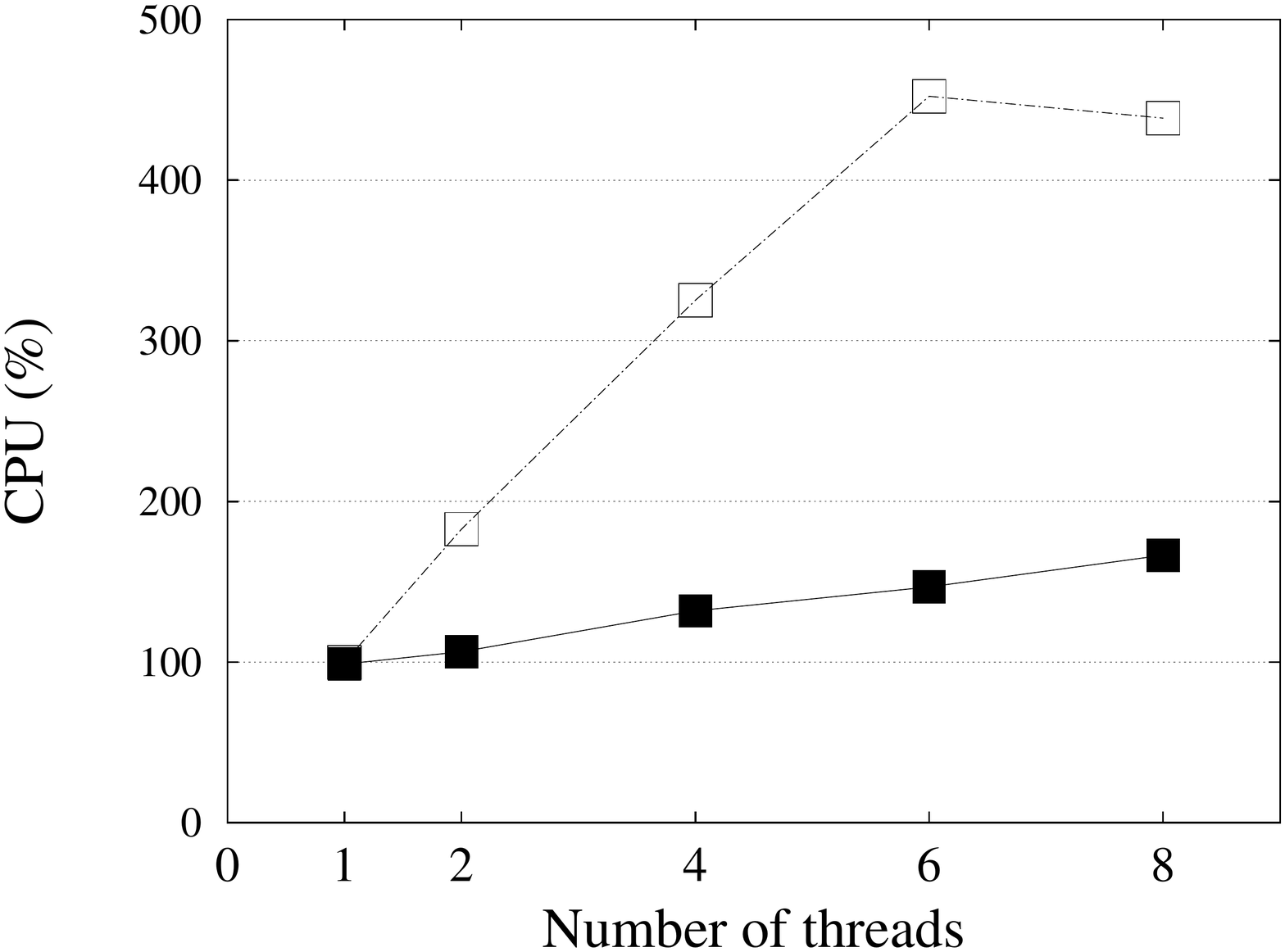}\\
       
    \end{tabular}
    \caption{The impact of the number of threads on the performance of P-SMR and opt-PSMR; the following metrics are shown: maximum throughput in Kilo commands executed per second (Kcps) (top-left); normalized per-thread throughput (bottom-left); fail rate and average latency in milliseconds (top-right); CPU usage (bottom-right)}
    \label{fig:scalability:writes}
 \end{center}
\end{figure*}

\section{Evaluation}
\label{sec:evaluation}

In this section, we describe the environment in which we conducted our experiments, comment on the implementation, explain the experimental setup and the rationale behind the experiments, and then report on our findings on the performance of opt-PSMR and how it relates to other techniques.

\subsection{Hardware setup}

We ran all the tests on a cluster with two types of nodes:
(a)~HP SE1102 nodes equipped with two quad-core Intel Xeon L5420 processors running at 2.5~GHz and 8~GB of main memory, and (b)~Dell SC1435 nodes equipped with two dual-core AMD Opteron processors running at 2.0~GHz and 4~GB of main memory. The HP nodes are connected to an HP ProCurve Switch 2910al-48G gigabit network switch, and the Dell nodes are connected to an HP ProCurve 2900-48G gigabit network switch. Each node is equipped with two network interfaces. The switches are interconnected via a 20~Gbps link. The nodes ran CentOS Linux 6.2 64-bit with kernel 2.6.32. Clients were deployed on the Dell nodes; Paxos's acceptors and servers were deployed on the HP nodes.


\subsection{Implementation}

We use a B$^+$-tree service to evaluate and compare the SMR, P-SMR, and opt-PSMR techniques. 
Each entry includes an 8-byte integer key, used as the tree index, and an 8-byte value. The service supports all the commands described in Section~\ref{sec:opsmr-motiv}. There are two replicas and the tree is initialized with 10 million keys on each replica. 

We implemented atomic multicast using Multi-Ring Paxos~\cite{MPP2012}.
Multicast groups in Multi-Ring Paxos are mapped to one or more Ring Paxos instances~\cite{MPSP2010}.
In Multi-Ring Paxos, a message can be addressed to a single group only, not to multiple groups.
To implement P-SMR and opt-PSMR, each server thread $t_i$ in our prototypes belongs to two groups: one group, $g_i$, to which no other thread in the server belongs, and one group $g_{all}$, to which every thread in each server belongs.

The safety check function of opt-PSMR for the B$^+$-tree is implemented as follows. 
The key space is range partitioned among the threads. 
Whenever a thread $t_i$ receives an \texttt{insert(k)} or \texttt{delete(k)} operation that is optimistically multicast, it first locates the leaf node $\alpha$ in the B$^+$-tree where the key will be inserted in or deleted from. 
Node $\alpha$'s parent in the tree points to $\alpha$ so that any key within range $\beta_1..\beta_2$ will be directed to $\alpha$.
The safety check passes if the following conditions hold:
(a)~the insertion or deletion of the key will not cause structural changes in the tree (e.g., the leaf node has space for the insert or will not result in a merged in the case of a delete);
(b)~the largest key in the partition assigned to thread $t_{(i-1)}$ is smaller than $\beta_1$; and
(c)~the smallest key in the partition assigned to thread $t_{(i+1)}$ is greater than $\beta_2$.
Note that conditions (b) and (c) check that no thread other than $t_i$ will be accessing node $\alpha$ during the insertion (or deletion).

\subsection{Experimental setup}

In all the experiments clients select the keys uniformly. Each experiment (i.e., a point in the graphs) is performed for 60 seconds out of which the first and last 5 seconds are discarded. We perform three sets of experiments: 


\begin{itemize}

\item The first experiment measures the cost of failed commands in opt-PSMR (see Section~\ref{exp:aborts}). Failed commands pass through the agreement layer twice and are expected to negatively impact  latency. 

\item The initial objective behind opt-PSMR is to overcome the inefficiency of P-SMR at executing dependent commands. In this experiment we vary the percentage of dependent commands in the workload and seek to see whether opt-PSMR achieves its goal in optimizing P-SMR (see Section~\ref{exp:mixed}). 

\item With a workload composed of dependent commands only, as the number of threads in P-SMR increases the performance reduces. In this experiment we compare the performance of opt-PSMR versus P-SMR while varying the number of threads (see Section~\ref{exp:scalability}). 

\end{itemize}

\subsection{The Impact of failed commands on performance}
\label{exp:aborts}

Figure~\ref{fig:aborts-latency} shows the effect of failed commands on the latency of opt-PSMR. 
There are 8 threads on each replica and the workload is composed of \texttt{insert} and \texttt{delete} operations only. 
Therefore, all the commands are optimistically multicast. 
Differently from the algorithm, in our implementation when a command fails, a replica notifies the corresponding client to resubmit the command. 
Thus, failed commands traverse the path between the client and the server twice. 
In the algorithm, as presented in Section~\ref{sec:alg}, the replicas multicast failed requests themselves without informing the clients.

The top left graph shows the fail rate versus the throughput. As the throughput increases the fail rate decreases. The reason for the decrease is that although the number of failed commands increases (bottom left graph), this growth is not proportional to the increase in throughput. The right graph shows three curves for latency: the average latency for all the commands, average latency for failed commands, and average latency for passed commands. As expected, the latency of failed commands is approximately twice the latency of passed commands. Since the number of failed commands is much lower than the number of passed commands, the impact of fails on the average latency is negligible.

\subsection{The impact of dependent commands on performance}
\label{exp:mixed}

Figure~\ref{fig:mixed-workloads} shows the maximum performance of SMR, P-SMR, and opt-PSMR with a workload composed of \texttt{read}, \texttt{insert}, and \texttt{delete} operations. Replicas of P-SMR and opt-PSMR contain 8 threads each. In P-SMR \texttt{insert} and \texttt{delete} operations are multicast to all the groups and therefore delivered by all the worker threads. In opt-PSMR however, these operations are optimistically multicast based on the keys they access. While the $x$-axis shows the percentage of dependent commands in P-SMR and SMR, it shows the percentage of optimistically multicast commands in opt-PSMR. 

As the percentage of dependent commands in the workload increases, the throughput of P-SMR decreases and at about 50\% falls below SMR's throughput. opt-PSMR's throughput however, remains above SMR's throughput even with 100\% of dependent commands in the workload (left graph). As the percentage of the dependent commands in the workload increases from 0 to 1, P-SMR suffers a big reduction in throughput. This is an evidence of P-SMR's inefficiency at executing dependent commands even when dependent commands constitute only a small fraction of the workload. opt-PSMR outperforms P-SMR by a factor of 2.4 times. 

The top graph on the right shows the fail rate in opt-PSMR. The low fail rate is the reason opt-PSMR outperforms SMR and P-SMR irrespective to the percentage of the dependent commands in the workload. Latency is shown in the bottom right graph. Latency of opt-PSMR is slightly higher than the latency of SMR and P-SMR, mainly because opt-PSMR achieves a higher throughput. 

\subsection{The impact of the number of threads on performance}
\label{exp:scalability}

Figure~\ref{fig:scalability:writes} shows the scalability of opt-PSMR and P-SMR with a workload composed of 100\% dependent commands. As the top left graph indicates, by adding more threads to the replicas, the throughput of opt-PSMR increases and the throughput of P-SMR decreases. Notice that the workload includes only dependent commands and thus all the threads in P-SMR must deliver these commands and synchronize to provide exclusive access of the service to only one of the threads. As more threads are added to P-SMR, the synchronization overhead increases and negatively affects the throughput. The low fail rate of opt-PSMR (top right graph) helps it achieve scalable performance.

The bottom left graph shows the normalized per-thread throughput. Although opt-PSMR does not reach perfect scalability due to the failed commands, its scalability is better than P-SMR's. The values of the latency in the top right graph show that opt-PSMR's gain in throughput does not incur high costs on latency. The bottom right graph shows the CPU usage of the techniques. opt-PSMR has higher CPU consumption due to the higher number of requests executed.


\section{Related Work}
\label{sec:rwork}
In Section~\ref{sec:survey} we have provided a thorough discussion on parallel state-machine replication and reviewed the related work, in this section we review general-purpose approaches that can be used to implement parallel replicas and then briefly overview optimistic approaches that are applied to replication techniques. 

\noindent \textbf{General-purpose approaches.} 
Allowing multiple threads to execute commands concurrently may result in state and output inconsistencies if dependent commands are scheduled differently in two or more replicas. In~\cite{AWHF2010,bhcls2010, DLCOM2009, TA2010} the authors propose different approaches to enforcing deterministic multithreaded execution of commands. These solutions impose performance overheads and may require re-development of the service using new abstractions. Another solution is to allow one of the multithreaded replicas to execute commands non-deterministically and log the execution path, which will be later replayed by the rest of the replicas. Logging and replaying have been mainly developed for debugging and security rather than fault tolerance~\cite{AS2009, DLFC2008, PC2008, PZXYKLL2009, RB1999, VLWOCFN2011, XBH2003}. These approaches typically have high overhead due to logging and may suffer from inaccurate replay, leading to differences among original and secondary copies. 

\noindent \textbf{Optimistic techniques.} Optimistic or speculative execution has been suggested before as a mechanism to reduce the latency of agreement problems. For example, in \cite{Kotla:2007, Wester:2009} clients are included in the execution of the protocol to reduce the latency of Byzantine fault-tolerant agreement. In \cite{JPPM02, KPAS99} the authors introduce atomic broadcast with optimistic delivery in the context of replicated databases. Similar to~\cite{marandi2011high} the motivation is to overlap the execution of transactions or commands with the ordering protocol, optimistically assuming that the outcome of the agreement layer will comply with the execution order. Our optimistic strategy differs from these approaches in that it only involves clients and replicas and not the agreement layer. Moreover for some applications, a safety check is sufficient to avoid the need for execution rollbacks, as we showed with a B$^+$-tree example. 


\section{Conclusion}
\label{sec:final}

State-machine replication is a well established replication technique and has been extensively discussed in the literature. 
In this paper, we concentrated on works that deal with adapting state-machine replication to parallel services. 
We reviewed existing proposals and compared their architectures. Our comparison showed that among existing techniques, P-SMR has a more scalable architecture in that unlike other approaches its design model does not include centralized components. We built on the scalable design of P-SMR and by identifying its shortcomings proposed  a novel technique based on an optimistic strategy that was able to significantly boost its performance.


\bibliographystyle{ieeetr}
\bibliography{main}

\end{document}